\documentclass[prc,aps,twocolumn,showpacs,amssymb,superscriptaddress,float]{revtex4}

\usepackage{amsmath}
\usepackage{graphicx}
\usepackage{dcolumn}
\usepackage{float}

\begin{document}

\title{Similarity of nuclear structure in $^{132}$Sn and  $^{208}$Pb regions:
proton-neutron multiplets}

\author{L. Coraggio$^{1}$, A. Covello$^{1,2}$, A. Gargano$^{1}$,
and N. Itaco$^{1,2}$} 
\affiliation{$^{1}$Istituto Nazionale di Fisica Nucleare, 
Complesso Universitario di Monte S. Angelo, I-80126 Napoli,
Italy\\
$^{2}$Dipartimento di Scienze Fisiche, Universit\`a
di Napoli Federico II,
Complesso Universitario di Monte S. Angelo,  I-80126 Napoli,
Italy}

\date{\today}

\begin{abstract}

Starting from the striking similarity of proton-neutron multiplets
in $^{134}$Sb and $^{210}$Bi, we perform a shell-model study of  
nuclei with two additional  protons or neutrons  to find out to
what extent this analogy persists.
We employ effective interactions derived  
from the CD-Bonn nucleon-nucleon potential  
renormalized by use of the $V_{\rm low-k}$ approach.
The calculated results for $^{136}$Sb, $^{212}$Bi, 
$^{136}$I, and $^{212}$At are in very good agreement with the available 
experimental data.
The similarity between $^{132}$Sn and  $^{208}$Pb regions is 
discussed in connection with the effective interaction, emphasizing
the role of core polarization effects. 

\end{abstract}    
\maketitle

Since its advent more than fifty  years ago, the shell model has been the 
basic framework for understanding the structure of complex nuclei in terms of 
individual nucleons. Within this model, nuclei around doubly closed shells play
a special role. In fact, they yield direct information on the two basic 
ingredients of the model: single-particle (SP) energies and matrix elements of 
the  effective interaction.
This makes them the best testing ground for realistic shell-model 
calculations where the effective interaction is derived from the free 
nucleon-nucleon ($NN$) potential.

For a long time our knowledge of  nuclei with few-valence particles or 
holes has been mostly limited to neighbors of  stable or long-lived 
doubly magic $^{16}$O, $^{40}$Ca,  $^{48}$Ca, $^{56}$Ni,  and $^{208}$Pb. 
However, during the last decade there has been substantial 
progress in the experimental study of  nuclei far from the stability line, and 
the development of radioactive nuclear beams is currently giving strong 
impetus to the study of exotic nuclei  around $^{78}$Ni, $^{100}$Sn,
and $^{132}$Sn.
These new data pose challenging  questions about the evolution 
of  the shell structure, as for instance the validity of magic numbers
when  moving far away from stability  and the existence of
possible changes in the mean field as  well as in the two-body  
interaction \cite{Dobaczewski07,Grawe07,Otsuka08}.

In this context, nuclei ``north-east'' of $^{132}$Sn are of special 
interest, since  in recent experiments some peculiar properties 
have been observed which  might be interpreted  
as the onset of a shell-structure modification 
(see introductory discussion in \cite{Coraggio06}).
To investigate whether these features really depend on the
exoticism of  $^{132}$Sn neighbors, we believe that a  comparative study of 
their spectroscopic properties and  those of nuclei close  to stable $^{208}$Pb
is desirable.

As is well known,  $^{132}$Sn and $^{208}$Pb, which  are
very differently  located  with respect to the  valley of stability,
both exhibit a  strong  neutron-proton asymmetry and strong shell 
closures. The existence of a  specific resemblance between  
$^{132}$Sn and $^{208}$Pb regions was 
pointed out long ago in Ref.~\cite{Blomqvist81}, where it was noticed  that
every SP proton or neutron state in the   $^{132}$Sn region, characterized  
by quantum numbers $(n\, l\, j)$, has its counterpart  around $^{208}$Pb
with quantum numbers $(n\, l+1\, j+1)$.  Based on this resemblance, 
the discrepancies  
between  experimental and Woods-Saxon SP  energies in  
$^{208}$Pb region  were then used to correct the SP energies  calculated  
for  $^{132}$Sn region, so as  to predict the energies of unobserved states. 
A few years later, an analysis similar in spirit to that of 
Ref.~\cite{Blomqvist81} was 
carried out in \cite{Leander84} using SP energies calculated with 
three different independent-particle models.  

Until recent years, however,  the data available for nuclei around $^{132}$Sn 
have  not been sufficient  to clearly assess  the similarity of the
spectroscopy of $^{132}$Sn and $^{208}$Pb  regions.
On the other hand, in several recent papers this similarity  has 
been exploited to interpret new observed levels in $^{132}$Sn 
neighbors (see for instance 
Refs.~\cite{Zhang96,Urban99,Fornal01,Korgul07,Isakov07}). 
In this regard, it is of key importance the fact that, given the 
correspondence between  the SP levels, the matrix elements of 
the effective interactions  in  $^{132}$Sn and  $^{208}$Pb regions are 
expected to be proportional  to one another.
This has stimulated several shell-model 
calculations on nuclei around $^{132}$Sn
\cite{Sarkar01,Sarkar04,Shergur05} with
two-body effective interactions originating from the modified version 
\cite{Warburton91} of the Kuo-Herling interaction \cite{Kuo71},
originally designed for the Pb region. 
However,  these attempts
have not been successful, as discussed for instance  in  \cite{Shergur05}, 
where the conclusion was drawn that a consistent Hamiltonian 
for the three nuclei beyond the $N=82$ shell closure,
$^{134}$Sb, $^{135}$Sb, and   $^{134}$Sn,
had  yet to be found.

In the works of Refs. \cite{Coraggio05,Coraggio06,Coraggio07a} we have shown 
that  the properties of these nuclei are well accounted for by a unique 
shell-model Hamiltonian with SP energies taken from experiment and two-body
effective interaction derived from the CD-Bonn $NN$ 
potential~ \cite{Machleidt01}. Along the same lines  we have  performed
a calculation for $^{210}$Bi, obtaining results in very good 
agreement with experiment \cite{Coraggio07}.

Based on these results, we have found it interesting to perform a 
comparative  shell-model study of  
$^{132}$Sn and $^{208}$Pb regions using  the same Hamiltonians as 
in our aforementioned studies. Here, we focus 
on odd-odd nuclei extending our calculations for
$^{134}$Sb and  $^{210}$Bi to systems with an  additional pair of protons
or neutrons. These are $^{136}$I and $^{136}$Sb in  $^{132}$Sn region and
their counterparts in $^{208}$Pb region, $^{212}$At and $^{212}$Bi.
A main aim of this study is to emphasize the striking similarity 
between proton-neutron multiplets in $^{134}$Sb and  $^{210}$Bi,
which is successfully reproduced by our effective interactions, and
investigate to what extent 
these interactions   predict persistence of similarity when adding  
two identical particles. It is worth pointing out that our effective
interaction virtually accounts for excitations left out from the chosen 
shell-model space. We shall see that a crucial role is played by the
renormalization induced by the core through one particle-one hole (1p1h)
excitations. 

In our shell-model calculations for $^{132}$Sn neighbors we assume that
the valence protons occupy the five levels $0g_{7/2}$, $1d_{5/2}$, $1d_{3/2}$,
$2s_{1/2}$, and  $0h_{11/2}$ of the 50-82 shell, while for neutrons the
model space includes the six levels $1f_{7/2}$, $2p_{3/2}$, $0h_{9/2}$, 
$2p_{1/2}$, $1f_{5/2}$, and   $0i_{13/2}$ of the 82-126 shell.
Similarly, for $^{208}$Pb  neighbors we take as model space for the valence 
protons the six levels of the 82-126 shell and let the valence neutrons occupy 
the seven levels $1g_{9/2}$, $0i_{11/2}$, $0j_{15/2}$,
$2d_{5/2}$, $3s_{1/2}$, $1g_{7/2}$,  and $2d_{3/2}$ of the 126-184 shell.   

As mentioned above, for both regions the two-body effective interaction is 
derived from the  CD-Bonn $NN$ potential. Details on the derivation, 
as well as on the adopted SP proton and neutron 
energies, can be found in Refs.~\cite{Coraggio05} and \cite{Coraggio07} for  
$^{132}$Sn and  $^{208}$Pb, respectively. We only mention here that the 
short-range repulsion of the $NN$ potential is renormalized by means of the 
$V_{\rm low-k}$ potential~\cite{Bogner02}, which is then used, with the
addition of the Coulomb force for protons,  to derive
the effective interaction $V_{\rm eff}$ within the framework of the
$\hat Q$-box folded-diagram expansion \cite{Coraggio09}. The calculation of 
the 
$\hat Q$-box is performed at second order in $V_{\rm low-k}$, which from now
 on denotes the renormalized nuclear plus Coulomb interaction. Namely,  
we include
four two-body terms: the $V_{\rm low-k}$, the two core polarization diagrams
$V_{\rm 1p1h}$ and $V_{\rm 2p2h}$, corresponding to one particle-one hole  and 
two particle-two hole excitations,  and a ladder diagram  accounting for 
excluded configurations above the chosen model space.
The  shell-model  calculations have been performed by using the NUSHELLX 
code \cite{NUSHELLX}.

\begin{figure}[H]
\begin{center}
\includegraphics[scale=0.45,angle=0]{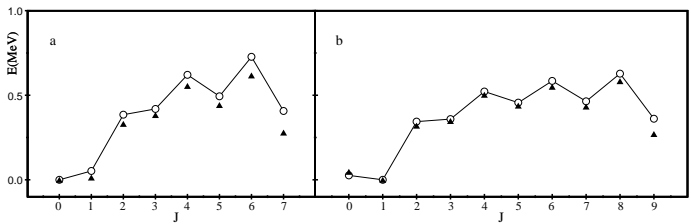}
\caption{(a) Proton-neutron $\pi g_{7/2} \nu f_{7/2}$ 
multiplet in $^{134}$Sb.(b) Proton-neutron $\pi h_{9/2} \nu g_{9/2}$ 
multiplet in $^{210}$Bi. The theoretical results are represented by open 
circles while the experimental data by solid triangles.}
\end{center}
\end{figure} 

To start with, we consider $^{134}$Sb and $^{210}$Bi. We 
report in Fig.~1 the calculated and experimental 
proton-neutron multiplets in  $^{134}$Sb arising from the 
$\pi g_{7/2} \nu  f_{7/2}$ configuration  together with  those 
in  $^{210}$Bi arising from the $\pi h_{9/2} \nu  g_{9/2}$ configuration, 
just as  they are shown  in  Refs.~\cite{Coraggio06} and \cite{Coraggio07},
respectively. This gives clear evidence of the  striking 
similarity  of the multiplets in the two nuclei.  
We see that in both cases a sizable energy gap exists between the $2^-$ state 
and 
the nearly degenerate $0^-$ and $1^-$ states, although the ground state
is $0^-$ in $^{134}$Sb and $1^-$ in $^{210}$Bi.
As discussed in detail in \cite{Coraggio07}, the measurement  of the 
ground-state spin in $^{210}$Bi as $1^-$  dates back to 
some 50 yeas ago~\cite{Erskine62} and since then the explanation of this 
peculiar feature has attracted great interest.  For $^{134}$Sb, instead,  
the observation of the  $0^-$ 
and $1^-$ states was only  a  recent experimental achievement 
~\cite{Shergur05,Korgul02}.
Note that our calculations  account for the position  of the  
$0^-$ and $1^-$ states in both nuclei.

A common feature of the multiplets  in $^{134}$Sb and 
$^{210}$Bi  is also a 
distinctive energy staggering  with the same magnitude and phase 
between the odd and even members starting from the  $3^-$  
state. As a consequence, the maximum aligned states come  down in energy 
becoming  the second exited states  in both nuclei, which makes them isomers.

From Fig. 1 we also see that our results are in very good agreement with 
experiment, as discussed in detail in  Refs.~\cite{Coraggio06} 
and \cite{Coraggio07}.
 This testifies to the soundness of our proton-neutron 
effective interactions,  in  particular  as regards
the diagonal two-body matrix elements for   the $\pi g_{7/2} \nu  f_{7/2}$ 
and $\pi h_{9/2} \nu  g_{9/2}$  configurations. The members of the multiplets
are in fact characterized by very little configuration mixing, the
percentage of the leading component ranging from 100 to 88 \% in $^{134}$Sb
and from  100 to 91 \% in $^{210}$Bi.

\begin{table}[H]
\caption{Diagonal matrix elements of $V_{\rm low-k}$ and 
$V_{\rm 1p1h}$ (in MeV) for proton-neutron configurations in $^{132}$Sn and $^{208}$Pb regions.} 
\begin{ruledtabular}
\begin{tabular}{lcccc}
& \multicolumn {2} {c} {$\pi 0g_{7/2} \nu 1f_{7/2}$} &
 \multicolumn {2} {c} { $\pi 0h_{9/2} \nu 1g_{9/2}$}  \\
 \cline{2-3} \cline{4-5}
$J^{\pi}$& $V_{\rm low-k}$ & $V_{\rm 1p1h}$ &
$V_{\rm low-k}$ & $V_{\rm 1p1h}$ \\
\colrule
$0^-$ &-0.596 &  0.089 & -0.468&   0.049\\
$1^-$ &-0.371 & -0.079 & -0.311 & -0.105\\
$2^-$ &-0.350 &  0.167 &  -0.292 &  0.115\\
$3^-$ &-0.210 &  0.047 & -0.165 &  0.011\\
$4^-$ &-0.165 &  0.156 & -0.146 &  0.115\\
$5^-$ &-0.197 &  0.079 & -0.131 &  0.045\\
$6^-$ &-0.070 &  0.146 & -0.079 &  0.099\\
$7^-$ &-0.348 &  0.080 & -0.151 &  0.059\\
$8^-$ &       &        & -0.035 &  0.101\\
$9^-$ &       &        & -0.274 &  0.059\\
\end{tabular}
\end{ruledtabular}
\end{table}

To have a better insight into the nature of our effective interactions, we  
have analyzed the various terms which 
contribute to them to find out their relative importance in  determining 
the final values of the matrix elements. In both  $^{132}$Sn and 
$^{208}$Pb regions, it turns out that the major contribution to produce 
the right  $0^{-}-1^{-}$  spacing  arises from virtual interactions with  
core particles  induced by the $NN$ potential, more precisely from
core polarization through 1p1h excitations. 
To evidence the role of core excitations, in Table~I we report
the diagonal matrix  elements of $V_{\rm low-k}$ and $V_{\rm 1p1h}$ for 
the $\pi g_{7/2} \nu  f_{7/2}$ and $\pi h_{9/2} \nu  g_{9/2}$  configurations.
From this table we see that the matrix elements  of  $V_{\rm low-k}$ as
well as   $V_{\rm 1p1h}$ have practically the same behavior and weight in the
two different mass regions. Note that the 1p1h contribution  substantially 
modifies in some cases the bare $V_{\rm low-k}$ interaction. In particular, it
is the 1p1h term that, having opposite sign for the $0^-$ and $1^-$ states, 
brings the latter down in energy.  
It is worth mentioning that not only the contribution from 
1p1h excitations but  also the $V_{\rm 2p2h}$ and ladder terms 
go in the same direction, as shown in~\cite{Coraggio06,Coraggio07}.  

\begin{figure}[H]
\begin{center}
\includegraphics[scale=0.45,angle=0]{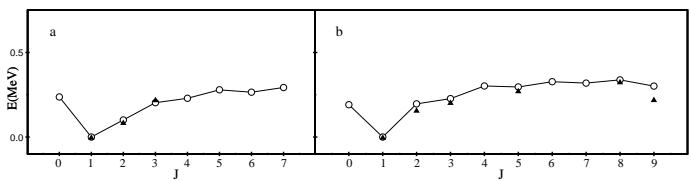}
\caption{Low-lying proton-neutron  
multiplet in  $^{136}$I (a) and $^{212}$At (b). 
The theoretical results are represented by open 
circles while the experimental data by solid triangle. See text for comments.}
\end{center}
\end{figure} 

We are now going to present our results for the   two counterpart pairs, 
$^{136}$I~-~$^{212}$At and $^{136}$Sb~-~$^{212}$Bi, with  two more 
valence protons and neutrons,  respectively. 
In Figs. 2 and 3, the calculated energies for the former and latter pair
are compared with the experimental ones \cite{NNDC,Urban06,Simpson07}.
We show in Fig. 2a the calculated energies of $^{136}$I for 
the lowest states 
with $J^\pi$=$0^-$ to $7^-$, which are dominated by the  
$(\pi g_{7/2})^{3} \nu  f_{7/2}$ configuration, and   in Fig. 2b
the energies of the $J^\pi$=$0^-$ to $9^-$  states in   $^{212}$At, 
arising  from the  $(\pi h_{9/2})^{3} \nu  g_{9/2}$ configuration.
In Figs. 3a and 3b, we report the energies of the corresponding states 
in  $^{136}$Sb and  $^{212}$Bi, which are dominated by 
the $\pi g_{7/2} (\nu  f_{7/2})^{3}$ and $\pi h_{9/2} (\nu  g_{9/2})^{3}$
configurations, respectively. Note that 
our predictions for $^{136}$Sb were first 
reported in the work of Ref.~\cite{Simpson07} to help interpreting some
new experimental data. 

\begin{figure}[H]
\begin{center}
\includegraphics[scale=0.45,angle=0]{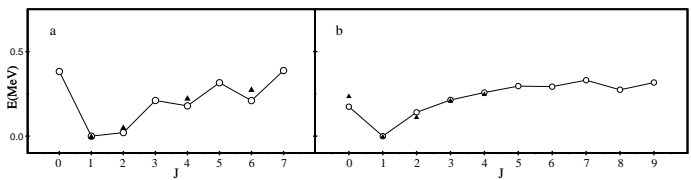}
\caption{Same as Fig.~(2) for $ ^{136}$Sb (a) and 
$^{212}$Bi (b).}
\end{center}
\end{figure} 

For all four nuclei, the considered states receive significant contributions 
from configurations other than the dominant one. In fact, we find that the 
the percentage of the $(\pi g_{7/2})^{3} \nu  f_{7/2}$ configuration for the  
states of  $^{136}$I and  of the $(\pi h_{9/2})^{3} \nu  g_{9/2}$
configuration for the states of $^{212}$At ranges from 64 to 75\%   and 
from 66 to 75\%, 
respectively. As for $^{136}$Sb and  $^{212}$Bi, it turns out
that the percentage of the $\pi g_{7/2} (\nu  f_{7/2})^{3}$ and 
$\pi h_{9/2} (\nu  g_{9/2})^{3}$ configurations goes from  58 to 76\% 
and from  48 to 75\%. Note that the configuration mixing is particularly
large for $^{212}$Bi,  the percentage of components other than the dominant one
exceeding 50\% for the 
$0^-$ state.  We shall come back to this point later.

As regards the experimental states, we have excluded those 
with no or multiple  spin-parity assignment. This is the case of the
$7^-$  state in $^{136}$I, whose position is still a matter of discussion 
\cite{Urban06,Fogelberg07}.
From Figs. 2 and 3 we see  that the agreement between theory and 
experiment is very 
good, the largest discrepancy being 
about 80 keV for the $9^-$ state in $^{212}$At.
Note, however, that some experimental levels are  still missing. 

Although  the observation of the missing states is certainly needed to 
further verify the outcome  of our calculations, 
the present comparison between theory and experiment encourages
use of our predictions to study how the structure of 
proton-neutron multiplets is affected when adding two identical particles.
From this point of view, the theoretical curves of 
Figs. 2 and 3  may be considered the evolution of the  
$\pi g_{7/2} \nu f_{7/2}$ and $\pi h_{9/2} \nu g_{9/2}$ multiplets.  

These curves have all a similar shape, but quite different from that 
of the multiplets   in  $^{134}$Sb and $^{210}$Bi. The main new
features are
an overall  flattening of the staggering and an energy  
increase of the  $0^-$ state with respect to the $1^-$ one. 
These are a manifestation of the dominant role played 
by pairing correlations, 
which means that the two additional protons or neutrons in both 
$A=136$ and 212
systems are prevalently coupled to zero angular momentum. However,
the wave functions of the states reported in Figs. 2 and 3 
contain also components with the pair coupled to $J \neq 0$, their 
weight being more significant when a neutron rather than a proton pair 
is added. As a matter of fact, it turns out that the proton-proton pairing
component of our effective interaction plays  a more significant role than 
that of the  neutron-neutron  one, which is largely due to the 
1p1h excitations.  
In Table II we report the diagonal matrix elements of $V_{\rm low-k}$ and 
$V_{\rm 1p1h}$ for the $(\pi 0g_{7/2})^2$, 
$(\pi 0h_{9/2})^2$,  $(\nu 1f_{7/2})^2$, and $(\nu 1g_{9/2})^2$ configurations.
Similarly to the proton-neutron case, we see that these proton-proton and 
neutron-neutron matrix elements have comparable magnitudes in 
$^{132}$Sn and $^{208}$Pb regions.
As already mentioned, in our 
calculation we explicitly include the  Coulomb force,
whose effect is clearly visible in columns 2 and 4 of Table II. 
The essential role of  1p1h excitations in determining the proton-proton 
pairing is evidenced by the large and negative values of the  $J^{\pi}= 0^+$ 
$V_{\rm 1p1h}$ matrix elements as compared to the $J \ne 0$ matrix elements.
As for the  $V_{\rm 1p1h}$ neutron-neutron matrix elements it is seen that 
they are all scaled down, which results in a a non negligible attenuation of 
the pairing force. 
This is borne out by the different experimental energies of the 
first excited 
states in the two-valence-neutron and -proton nuclei, the energy 
gap between the ground and the first $2^+$ states in $^{134}$Sn and $^{210}$Pb 
being  about 500 keV smaller than that in $^{134}$Te and $^{210}$Po.
As emphasized in  Ref.~\cite{Covello09}, our  shell-model study  
accounts for the  close resemblance between the nuclei 
of each of these  two pairs as well as for the difference between the  
two-valence-neutron and -proton nuclei. 

\begin{table}[H]
\caption{Diagonal matrix elements of $V_{\rm low-k}$ and 
$V_{\rm 1p1h}$ (in MeV) for proton-proton and neutron-neutron  configurations in $^{132}$Sn and $^{208}$Pb regions.} 
\begin{ruledtabular}
\begin{tabular}{lcccccccc}
& \multicolumn {2} {c} {$(\pi 0g_{7/2})^2$}  & 
 \multicolumn {2} {c} {$(\pi 0h_{9/2})^2$}&
\multicolumn {2} {c} {$(\nu 1f_{7/2})^2$} & 
\multicolumn {2} {c} {$(\nu 1g_{9/2})^2$}\\
 \cline{2-3} \cline{4-5} \cline{6-7} \cline{8-9}  
$J^{\pi}$&  $V_{\rm low-k}$ & $V_{\rm 1p1h}$& 
$V_{\rm low-k}$ & $V_{\rm 1p1h}$& $V_{\rm low-k}$ & $V_{\rm 1p1h}$& $V_{\rm low-k}$ & $V_{\rm 1p1h}$ \\
\colrule
$0^+$ &  0.063& -0.549&  0.079& -0.535& -0.403& -0.100& -0.227& -0.162\\
$2^+$ & -0.016&  0.071& -0.011& -0.013& -0.289&  0.018& -0.226& -0.017\\
$4^+$ &  0.124&  0.176&  0.099&  0.090& -0.136&  0.051& -0.116&  0.028\\
$6^+$ &  0.214&  0.237&  0.148&  0.114& -0.063&  0.067& -0.066&  0.045\\
$8^+$ &       &       &  0.199&  0.156&       &       & -0.029&  0.058\\
\end{tabular}
\end{ruledtabular}
\end{table}

As regards the effects of components with the pair 
coupled to $J \neq 0$, we have verified that their presence tends to produce 
staggering with phase opposite to that observed in  one proton-one  neutron 
systems.  
This is evident for $^{136}$Sb, but not  for $^{212}$Bi. 
The explanation lies in the fact that in $^{212}$Bi this effect is 
quenched by configuration mixing, which we have found to be more relevant 
for this nucleus.
As regards $^{136}$I and   $^{212}$At, 
we don't find evidence of significant staggering, the curves of Fig. 2 
being almost 
flat but for the lowest angular momenta.
This is because, as mentioned above, the wave functions of the 
considered states in 
these nuclei contain a small percentage of nonzero-coupled proton pair 
components,
giving rise to a staggering which is largely washed out by the configuration 
mixing. 

In summary, we have performed here a comparative shell-model study of 
proton-neutron multiplets in  $^{132}$Sn and $^{208}$Pb regions, focusing
on  the three far from stability nuclei  
$^{134}$Sb,  $^{136}$I, 
 $^{136}$Sb and their counterparts around stable $^{208}$Pb.
We have shown that the calculated energies are in very good 
agreement with the available experimental data for all the six nuclei 
considered and emphasized that a close resemblance between the spectroscopy 
of the two regions persists when moving away from one proton-one neutron 
systems.
A main achievement of our work is that this similarity emerges quite 
naturally from our shell-model interactions  which are derived from a 
realistic  $NN$ potential without any adjustable parameters.

We are confident that this work may stimulate new experiments in  
both $^{132}$Sn and $^{208}$Pb regions and be a guide to the 
interpretation of data.

\clearpage

\end{document}